\documentclass[%
reprint,
superscriptaddress,
showpacs,preprintnumbers,
aps,
prl,
]{revtex4-1}

\usepackage{amsmath}
\usepackage{bm}
\usepackage{mathrsfs}
\usepackage[hyperfootnotes=false]{hyperref} 
\usepackage{mathtools} 
\usepackage{braket}
\usepackage{amssymb}
\usepackage{dcolumn}
\usepackage[dvipdfmx]{graphicx, color}
\usepackage{color}
\usepackage[multiple]{footmisc} 

\usepackage[normalem]{ulem} 
\usepackage{xcolor,cancel} 

\DeclareMathOperator{\Tr}{Tr}

\newcommand{\Apeak}{A_{\mathrm{peak}}}

\begin{document}
\title{Time-dependent Hartree-Fock study of electron-hole interaction effects on high-harmonic generation from periodic crystals}

\author{Takuya Ikemachi}
  \email[]{ikemachi@gono.phys.s.u-tokyo.ac.jp}
  \affiliation{Department of Physics, Graduate School of Science, The University of Tokyo, 7-3-1 Hongo, Bunkyo-ku, Tokyo 113-0033, Japan}
\author{Yasushi Shinohara}
  \affiliation{Photon Science Center, Graduate School of Engineering, The University of Tokyo, 7-3-1 Hongo, Bunkyo-ku, Tokyo 113-8656, Japan}
\author{Takeshi Sato}
  \affiliation{Photon Science Center, Graduate School of Engineering, The University of Tokyo, 7-3-1 Hongo, Bunkyo-ku, Tokyo 113-8656, Japan}
  \affiliation{Department of Nuclear Engineering and Management, Graduate School of Engineering, The University of Tokyo, 7-3-1 Hongo, Bunkyo-ku, Tokyo 113-8656, Japan}
\author{\\Junji Yumoto}
  \affiliation{Department of Physics, Graduate School of Science, The University of Tokyo, 7-3-1 Hongo, Bunkyo-ku, Tokyo 113-0033, Japan}
  \affiliation{Institute for Photon Science and Technology, Graduate School of Science, The University of Tokyo, 7-3-1 Hongo, Bunkyo-ku, Tokyo, 113-0033 Japan}
  \affiliation{Research Institute for Photon Science and Laser Technology, The University of Tokyo, 7-3-1 Hongo, Bunkyo-ku, Tokyo, 113-0033 Japan}
\author{Makoto Kuwata-Gonokami}
  \affiliation{Department of Physics, Graduate School of Science, The University of Tokyo, 7-3-1 Hongo, Bunkyo-ku, Tokyo 113-0033, Japan}
\author{Kenichi L. Ishikawa}
  \affiliation{Photon Science Center, Graduate School of Engineering, The University of Tokyo, 7-3-1 Hongo, Bunkyo-ku, Tokyo 113-8656, Japan}
  \affiliation{Department of Nuclear Engineering and Management, Graduate School of Engineering, The University of Tokyo, 7-3-1 Hongo, Bunkyo-ku, Tokyo 113-8656, Japan}

\begin{abstract}
 We investigate the multielectron effects on high-harmonic generation from solid-state materials using the time-dependent Hartree-Fock theory.
 We find qualitative change in harmonic spectra, in particular, multiple-plateau formation at significantly lower laser intensities than within the independent-electron approximation.
 We reveal its origin in terms of interband polarization, i.e, electron-hole polarization, enabling interband excitation at remote crystal momenta via Coulomb potential.
\end{abstract}

\maketitle

Advances in ultrashort intense laser technique have given access to field-induced extreme nonlinear physics \cite{Corkum2007,Krausz2009}.
In particular, high-harmonic generation (HHG) from gas-phase materials has led to successful applications such as attosecond-pulse generation \cite{Goulielmakis2008,Zhao_2012} and coherent soft x-ray sources in the water-window \cite{Spielmann661,PhysRevLett.101.253901} and even in the keV region \cite{Popmintchev2012}, giving birth to attosecond science.
Many features of gas-phase HHG can be intuitively and quantitatively explained by a semiclassical three-step model \cite{Corkum1993,Kulander1993}.

Recently, solid-state materials have emerged as a new stage for strong-field physics, and in particular, many experimental observations of HHG from solids have been reported since the first discovery by Ghimire {\it et al.} \cite{Ghimire2011,Schubert_2014,Luu2015a,Vampa2015a,Hohenleutner2015a,Ndabashimiye_2016,Han2016,Garg2016,Liu2017,You2017,LangerF.2017,You2017a}.
These studies have revealed unique aspects of solid-state HHG such as linear scaling of cutoff energy with field strength \cite{Ghimire2011,Luu2015a} and multiple plateau structure \cite{Ndabashimiye_2016,You2017a}.
In contrast to gas-phase HHG, though, the mechanism underlying these features is still under intensive discussion.
HHG from solid-state materials have first been discussed in terms of Bloch oscillation, or the intraband current \cite{Ghimire2011,Hawkins2013,Schubert_2014,Luu2015a}.
More recently, it has been shown that the interband current makes a dominant contribution over the intraband current to the radiation above the band gap energy, while both contribute equivalently to the below-gap radiation \cite{Vampa2014,Wu2015,Otobe2016}.
In this context, Vampa {\it et al.} have proposed a real-space three-step model analogous to its gas-phase counterpart \cite{Vampa2015} to explain solid-state HHG in terms of interband current.
Higuchi {\it et al.} have proposed another real-space picture using localized Wannier-Stark (WS) states and the strong-field approximation, where the differences of quasienergies of WS states determine the radiation energies \cite{Higuchi2014a}.
We have recently proposed a solid-state momentum-space three-step model that considers electron dynamics across multiple bands, incorporating field-induced intraband displacement, interband tunneling, and recombination with the valence-band (VB) hole \cite{Ikemachi2017} (see also \cite{Wu2016,Du2017}).
With regard to numerical methods, the time-dependent Schr\"{o}dinger equation (TDSE) \cite{Korbman2013,Hawkins2015a,Wu2015,Guan2016,Ikemachi2017} and semiconductor Bloch equations (SBEs) \cite{Golde2008,Vampa2014,Hohenleutner2015a,Garg2016,LangerF.2017} have often been used, while some authors have tried {\it ab-initio} approaches based on the time-dependent density functional theory (TDDFT) \cite{Otobe2012,Otobe2016,Tancogne-Dejean2017}.

While most of these previous works \cite{Ghimire2011,Hawkins2013,Schubert_2014,Luu2015a,Vampa2014,Vampa2015a,Golde2008,Hohenleutner2015a,LangerF.2017,Ikemachi2017,Wu2016,Du2017} have used independent-electron approximation,
the role of the electron-hole interaction (EHI) in the strong field regime is largely unexplored.
EHI forms a characteristic resonance, i.e., excitons, in the linear response regime, which is well described by the {\it ab initio} Bethe-Salpeter equation \cite{Onida2002} based on many-body perturbation theory (MBPT).
It is, however, still a formidable task to describe highly nonlinear dynamics of extended systems within the framework of MBPT except for a few pioneering works \cite{Leitsmann2005,Attaccalite2011}.
Garg {\it et al.} have recently suggested that EHI affects harmonic yields from silicon dioxide using SBE incorporating the interelectronic interaction \cite{Garg2016}.

In this Letter, we study the effects of the electron-hole interaction on solid-state HHG using the time-dependent Hartree-Fock (TDHF) calculation.
Our results for a one-dimensional (1D) model crystal show that EHI qualitatively modifies harmonic spectra.
Especially, the second plateau appears at significantly lower laser intensity than within the independent-electron approximation.
In order to uncover its origin, we expand the TDHF equations with Houston basis \cite{Krieger1986} and reveal that the Coulomb interaction from the interband polarization at the minimum band gap, once formed, mediates excitation of distant VB electrons.
This mechanism is supported by the time-frequency structure of HHG and band populations.
The present study will pave the way toward the ultimate goal of revealing correlations in ultrafast electron dynamics in solids.

We solve a set of the spin-restricted TDHF equations in the velocity gauge, for an electron orbital $\psi_{b {\bm k}_0}$ that {\it initially} lies in band $b$ with crystal momentum ${\bm k}_0$ (atomic units are used throughout unless otherwise mentioned),
\begin{align}
 i \frac{\partial}{\partial t} &\psi_{b{\bm k}_0}({\bm x}, t) = \hat{h}(t) \psi_{b{\bm k}_0}({\bm x}, t) \nonumber \\
 &= \left[ [\hat{\bm p} + {\bm A}(t)]^2 / 2 + U({\bm x}) + \hat{w}[\rho(t)]\right] \psi_{b{\bm k}_0}({\bm x}, t),\label{eq:TDHF equation}
\end{align}
where ${\bm A}(t) = - \int_{0}^t {\bm E}(t^{\prime}) dt^{\prime}$ denotes the vector potential of the electric field ${\bm E}(t)$, $U({\bm x})$ the periodic potential from the crystal nuclei, $\rho(t)$ the density matrix,
\begin{equation}
 \rho ({\bm x}, {\bm x}^{\prime}, t) = 2\sum_{b\in {\rm VB}, \; {{\bm k}_0}} \psi_{b{\bm k}_0}({\bm x},t)\, \psi_{b{\bm k}_0}({\bm x}^{\prime},t)^*, \label{eq:density matrix operator}
\end{equation}
and the operator $\hat{w}[\rho]$ describes the contribution from the interelectronic Coulomb interactions, composed of the Coulomb and exchange terms (see Supplementary Information for details).
$\psi_{b{\bm k}_0}$(t) is initially the VB Bloch function $\phi_{b{\bm k}_0}$, obtained as the self-consistent eigenstate of the field-free Hartree-Fock Hamiltonian
\begin{equation}
 \hat{h}_0 = \hat{\bm p}^2 / 2 + U({\bm x}) + \hat{w}[\rho(0)]
\end{equation}
with the energy eigenvalue $\varepsilon_{b{{\bm k}_0}}$.
We calculate the HHG spectrum as the modulus square of the Fourier transform of the induced current ${\bm j}(t) = 2\sum_{b\in{\rm VB}, \; {{\bm k}_0}} \bra{\psi_{b{\bm k}_0}(t)} \hat{\bm p} + {\bm A}(t) \ket{\psi_{b{\bm k}_0}(t)}$.
It should be remembered that $b$ and ${\bm k}_0$ are the band index and crystal momentum, respectively, of the initial state.

In parallel, we also perform simulations without EHI using the {\it frozen} TDHF Hamiltonian
\begin{equation}
 \hat{h}_f(t) = [\hat{\bm p} + {\bm A}(t)]^2 / 2 + U({\bm x}) + \hat{w}[\rho_0],
\end{equation}
with $\rho_0({\bm x}, {\bm x}^{\prime}) = e^{-i{\bm A}(t) \cdot {\bm x}} \rho(0) e^{i{\bm A}(t) \cdot {\bm x'}}$, where electrons move independently in the potential constructed by the ground state Bloch functions.
The factors $e^{-i{\bm A}(t)\cdot {\bm x}}$ and $e^{i{\bm A}(t) \cdot {\bm x}'}$ are introduced since we use the velocity gauge.
Note that the {\it full} TDHF Hamiltonian can be written as $\hat{h}(t) = \hat{h}_f(t) + \hat{w}[\delta\hat{\rho}(t)]$, whose second term corresponds to EHI, with $\delta\hat{\rho}(t) = \hat{\rho}(t) - \hat{\rho}_0$.

\begin{figure}[tb]
 \centering
 \includegraphics[width=1.0\linewidth]{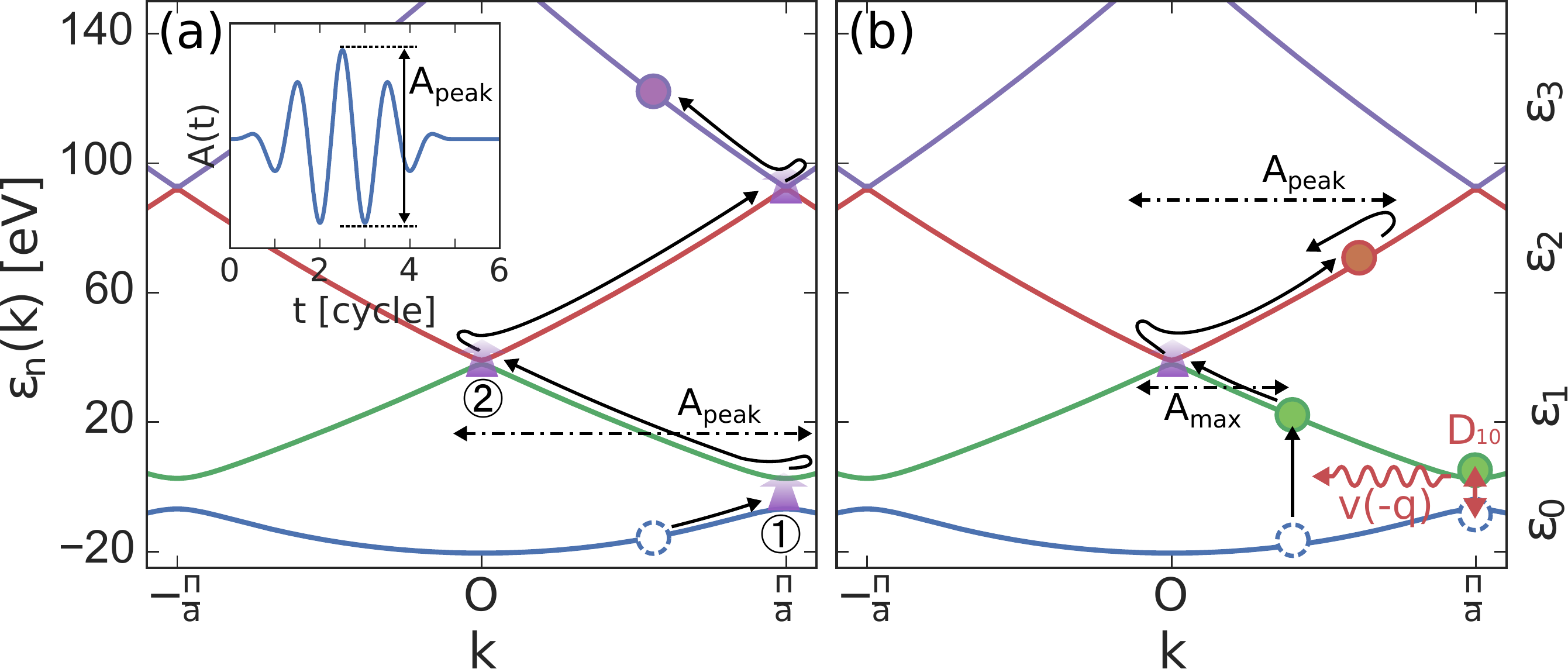}
 \caption{
 Pictorial representation of momentum-space electron dynamics (a) within the independent-electron approximation and (b) involving hauling-up excitation.
 The inset in (a) shows the waveform of the vector potential used in TDHF and {\it frozen} TDHF simulations and the definition of $\Apeak$.
 The single VB and first three CBs are shown for a 1D model hydrogen chain insulator (see text).
 The band index $n$ is labeled as $0, 1, 2, \dots$ from the bottom.
 }
 \label{fig:pictorical}
\end{figure}

We consider a 1D model crystal along laser polarization.
1D models have previously been used in several works \cite{Korbman2013,Higuchi2014a,Hawkins2015a,Wu2015,Ikemachi2017} to study the fundamental nature of solid-state HHG and turned out to be useful.
Moreover, a 1D system has a strong electron-hole correlation \cite{haug2009quantum}, thus, which is suitable for the investigation of EHI.
Specifically, our system is a 1D model hydrogen chain insulator with a lattice constant of $a = 3.6$, composed of a series of hydrogen dimers whose bond length is $1.6$.
We use a soft-Coulomb potential $v(x, x^{\prime}) = [(x-x^{\prime})^2 + 1]^{-1/2}$ for both electron-nucleus and electron-electron interactions.
Fig.~\ref{fig:pictorical} shows the band structure, the set of the energy eigenvalues $\varepsilon_{b{\bm k}_0}$, with a gap energy of $9.5$ eV.
The lowest band or VB is initially fully occupied.
Then we numerically integrate the TDHF equations (\ref{eq:TDHF equation}) and its counterpart with $\hat{h}_f(t)$ for a laser field $E(t) = E_0\sin^2(t/\tau)\sin(\omega t)$ with $\tau = 702.3 $ (5 cycle), $\hbar \omega = 0.387$ (eV), where $E_0$ denotes the field amplitude.

\begin{figure}[t]
 \centering
 \includegraphics[width=1.0\linewidth]{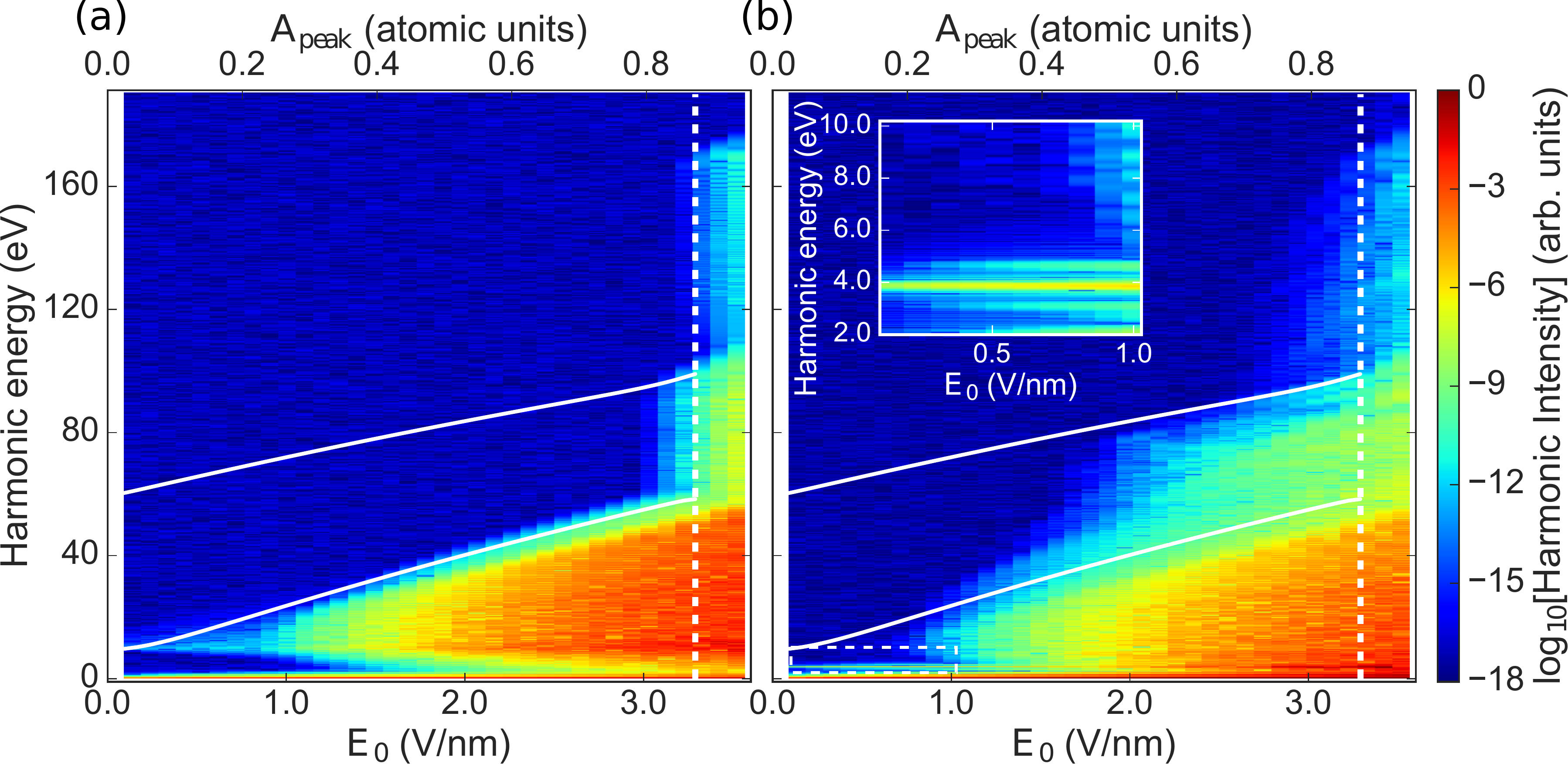}
 \caption{
 Harmonic spectra as functions of the field amplitude $E_0$ (bottom axis) and corresponding $\Apeak$ (top axis) obtained from (a) frozen TDHF and (b) full TDHF simulations.
 The white dashed vertical lines denote $\Apeak = \frac{\pi}{a}=0.87$, which characterizes the position where the multiple plateaus appear according to the solid-state three-step model \cite{Ikemachi2017}.
 Two white solid lines are the energy differences between CBs and VB as function of $\Apeak$, i.e., $\varepsilon_{10}(\frac{\pi}{a} - \Apeak)$ (lower) and $\varepsilon_{20}(\Apeak)$ (higher).
 Inset: close-up of the low-field region represented by a dashed rectangle in (b).
 }
 \label{fig:harmonic spectra}
\end{figure}

\begin{figure}[t]
 \centering
 \includegraphics[width=0.98\linewidth]{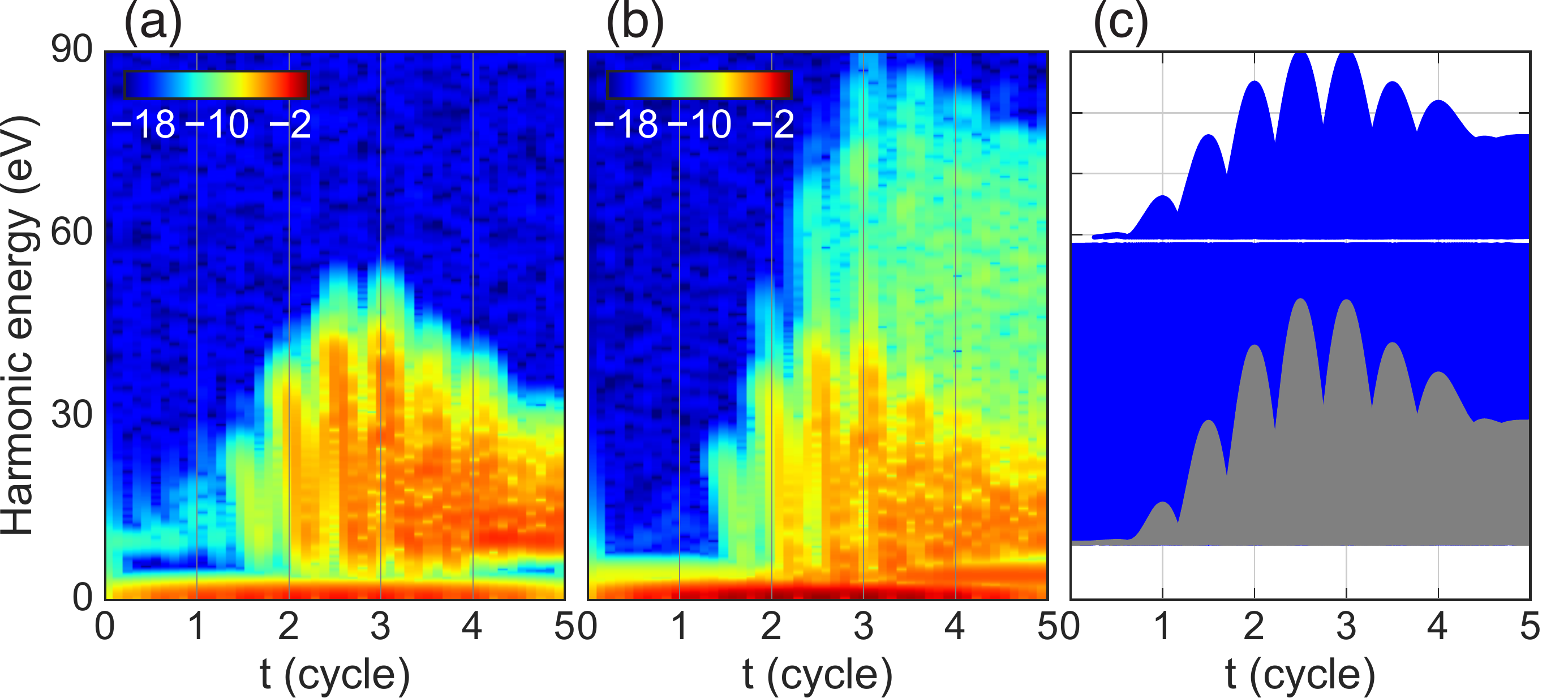}
 \caption{
 (a, b) Time frequency structure of HHG extracted by Gabor transform from (a) frozen TDHF and (b) full TDHF simulation results for $E_0 = 2.60$ V/nm. 
 (c) gray lines: electron trajectories that first tunnel from VB to the first CB band at the minimum band gap $k = \pm \frac{\pi}{a}$, drawn based on the solid-state three-step model \cite{Ikemachi2017}. 
 Blue lines: trajectories involving promotion to the first CB via hauling-up excitation.
 The blank at $\sim 60$ eV, looking like a white line, reflects the energy gap between the first and second CB.
 }
 \label{fig:time-frequency}
\end{figure}

The obtained harmonic spectra are shown in Fig.~\ref{fig:harmonic spectra} as the functions of the field amplitude $E_0$ and corresponding $\Apeak$, the maximum peak-to-valley amplitude of the vector potential $A(t)$ [see the inset of Fig.~\ref{fig:pictorical}(a)].
In the case of the {\it frozen} TDHF [Fig.~\ref{fig:harmonic spectra}(a)], i.e., within the independent-electron approximation, the appearance of multiple plateaus at $\Apeak = \frac{\pi}{a} = 0.87$ and the cutoff positions can be understood on the basis of the solid-state momentum-space three-step model \cite{Ikemachi2017,Wu2016,Du2017}.
Analogously to the case of the gas-phase three-step model \cite{Corkum1993,Kulander1993}, one can easily deduce many aspects of solid-state HHG by tracing electron dynamics in momentum space across multiple bands.
A typical trajectory is depicted in Fig.~\ref{fig:pictorical}(a).
An electron initially in the VB undergoes intraband displacement and gets excited at the minimum band gap (MBG) $k = \pm\frac{\pi}{a}$ ($\textcircled{\footnotesize 1}$) to the first conduction band (CB), say, at $t = t_0$.
The subsequent momentum displacement in the first CB is given by $A(t) - A(t_0)$, where $|A(t) - A(t_0)|$ is bounded by $\Apeak$.
Hence, if $\Apeak < \frac{\pi}{a}$, no excited electrons can reach the next MBG ($k=0$), and they only oscillate in the first CB, which forms a single plateau in the high-harmonic spectra.
On the other hand, if $\Apeak > \frac{\pi}{a}$, a part of electrons can reach the next MBG ($\textcircled{\footnotesize 2}$), be promoted to the second CB, and further climb up to higher and higher CBs by repeating the intraband displacement and interband tunneling, leading to the formation of multiple plateaus.
The cutoff positions as well as time-frequency structure of HHG (see Fig.~\ref{fig:time-frequency}) can be deduced by tracing all the trajectories starting from different initial crystal momenta $k_0$.
In Fig.~\ref{fig:harmonic spectra}(a) the second plateau appears slightly before $\frac{\pi}{a}$, because tunneling from the VB to CB takes place not only precisely at MBG but also in its vicinity.

Let us now turn on EHI, for which the full TDHF results are shown in Fig.~\ref{fig:harmonic spectra}(b).
We first notice an exciton peak at $3.8$ eV below the gap energy at low intensity [inset in Fig.~\ref{fig:harmonic spectra}(b)], which indicates that the TDHF simulations capture EHI appropriately (see Supplementary Information for the exciton energy and linear response).
Note that TDDFT at present cannot reproduce excitons, which is based on the simple adiabatic local-density approximation in practical implementations \cite{Onida2002}.
More remarkably, the second plateau already appears at $\Apeak \sim 0.5$, much smaller than $\frac{\pi}{a}$.
This is a striking manifestation of the EHI, which qualitatively alters HHG spectra.

In order to reveal the underlying microscopic mechanism, we expand the orbital functions $\psi_{b k_0}(x,t)$ with Houston states $e^{-i A(t) x} \phi_{nk(t)}(x)$ \cite{Krieger1986}, the instantaneous eigenstates of $\hat{h}_f(t)$ with eigenvalues $\varepsilon_{nk(t)}$, as
\begin{equation}
 \psi_{bk_0}(x,t) = \sum_m \alpha_{bk_0}^{m}(t) e^{-i \int_0^t \varepsilon_{m k(t^{\prime})} dt^{\prime}}e^{-i A(t) x} \phi_{mk(t)}(x),\label{eq:Houston function expansion of orbital}
\end{equation}
where $k(t) = k_0 + A(t)$ is the instantaneous crystal momentum incorporating {\it intraband} dynamics.
Since the system under consideration has a single VB, we drop the initial band index $b$ hereafter.
Substituting Eq.~(\ref{eq:Houston function expansion of orbital}) into Eq.~(\ref{eq:TDHF equation}) (see Supplementary Information for details), we obtain coupled equations for complex amplitudes $\alpha_{k_0}^m(t)$ expressing {\it interband} dynamics,
\begin{align}
 i&\frac{d}{dt} \alpha_{k_0}^m(t) = \sum_{n} \alpha_{k_0}^{n}(t) e^{i\int_0^t \varepsilon_{mn}[k(t^{\prime})]dt^{\prime}} \nonumber \\
 & \times \left[ E(t) d_{k(t)}^{mn} - \sum_{q \in {\rm BZ}} \bar{v}(-q) D_{k(t)+q}^{mn}(t) \right],\label{eq:EOM for Houston function expansion}
\end{align}
where $\varepsilon_{mn}(k) = \varepsilon_{mk} - \varepsilon_{nk}$ is the energy difference between band $m$ and $n$ at crystal momentum $k$, $\bar{v}(q)$ is the spatial Fourier transform of the interelectronic soft Coulomb potential, and $d_{k}^{mn} = i\braket{u_{km} | \nabla_k u_{kn}}$, with $u_{km}(x)$ is the lattice periodic part of the initial Bloch state, or $\phi_{km}(x) = e^{i k x} u_{km}(x)$.
$D^{mn}_{k(t)}$ denotes the time-dependent interband polarization between $m$ and $n$ at $k(t)$:
\begin{equation}
 D^{mn}_{k(t)}(t) = \alpha_{k_0}^m(t) \alpha_{k_0}^{n\ast}(t) e^{-i \int_0^t \varepsilon_{mn}[k(t^{\prime})] dt^{\prime}}.
\end{equation}

Assuming that population transfers from the VB to CBs are small (see Fig.~\ref{fig:electron final population} below), we introduce approximations $\alpha_{k_0}^{0}(t) \approx 1$ and $\alpha_{k_0}^{m \ge 1}(t) \approx 0$ \cite{McDonald2017}.
Then Eq.~(\ref{eq:EOM for Houston function expansion}) for the first CB ($m=1$) becomes
\begin{equation}
 i\frac{d}{dt} \alpha_{k_0}^1(t) \approx e^{i\int_0^t \varepsilon_{10}[k(t^{\prime})]dt^{\prime}} 
   \left[ E(t) d_{k(t)}^{10} - \sum_q \bar{v}(-q) D_{k(t)+q}^{10}(t) \right],
   \label{eq:EOM for Houston with FVB}
\end{equation}
which describes the excitation dynamics of a VB electron starting from crystal momentum $k_0$.

The first term of Eqs.~(\ref{eq:EOM for Houston function expansion}) and (\ref{eq:EOM for Houston with FVB}) comes from the {\it frozen} TDHF Hamiltonian, and thus describes the independent electron dynamics, depicted by the semiclassical trajectory analysis \cite{Ikemachi2017}.
The second term, on the other hand, stems from EHI $\hat{w}[\delta\rho(t)]$ [see Eq.~(S4) of Supplementary Information] and indicates that interband or electron-hole polarization at a remote crystal momentum $k(t) + q$,
\begin{equation}
 D_{k(t)+q}^{10}(t) = \alpha_{k_0+q}^1(t) e^{-i\int_0^t \varepsilon_{10}[k(t^{\prime})+q]dt^{\prime}},
\end{equation}
can induce quasi-resonant excitation when $\varepsilon_{10}[k(t)] \approx \varepsilon_{10}[k(t)+q]$.
Therefore, even if a VB electron starting from $k_0$ dose not reach MBG through intraband displacement, it can be excited to the first CB once another electron initially at $k_0+q$ reaches MBG and tunnels to the CB [Fig.~\ref{fig:pictorical}(b)].
It should be noticed that neither the first nor second terms directly change the crystal momentum, thus, the instantaneous crystal momentum is always given by $k(t) = k_0 + A(t)$, in whichever band the electron actually is. 

This {\it hauling-up} effect provides a shortcut for VB electrons to climb up to the second CB, which leads to the formation of the second plateau even if $\Apeak < \frac{\pi}{a}$.
The electrons initially at $k_0 \in [-\max(A(t)), -\min(A(t))]$ pass by $k=0$, i.e., MBG between the first and second CB.
Thus, if these VB electrons are excited to the first CB via the hauling-up effect, then they can climb up to the second CB by tunneling at $k=0$, eventually forming the second plateau via recombination with the VB hole.
Note that they cannot reach MBG at $k=\pm\frac{\pi}{a}$ between the second and third CB.
Therefore, the cutoff energy is expected to be given by $\varepsilon_{20}(\Apeak)$.
This prediction is in good agreement with the cutoff energy obtained from the TDHF simulation at $0.5 \lesssim \Apeak \le \frac{\pi}{a}=0.87$ [the upper white line in Fig.~\ref{fig:harmonic spectra}(b)].

Figure \ref{fig:time-frequency}(a) and (b) show the time-frequency structure of HHG extracted by Gabor transformation from the frozen TDHF and TDHF simulations, respectively, for field amplitude $E_0=2.60$ V/nm corresponding to $\Apeak = 0.69$.
For the frozen TDHF case, we can well reproduce the spectrogram by drawing momentum-space semiclassical trajectories, assuming tunneling at MBG, intraband displacement, and photoemission on recombination with the VB hole [gray lines in Fig.~\ref{fig:time-frequency}(c)] \cite{Ikemachi2017}.
The time-frequency structure from the full TDHF is significantly different, with photoemission above $60$ eV, but can be reproduced if we additionally consider vertical excitation to the first CB from VB at arbitrary moments via the hauling-up [blue lines in Fig.~\ref{fig:time-frequency}(c)].
We see some discrepancy in the early stage of the pulse; harmonics from the second CB ($\gtrsim 60$ eV) are observed only after second cycles in Fig.~\ref{fig:time-frequency} (b).
This indeed supports our view that hauling-up becomes effective only after sufficient interband polarization is formed at MBG.

To further verify the hauling-up mechanism by EHI, we compare, in Fig.~\ref{fig:electron final population}, the final band population obtained through projection onto the ground-state Bloch orbitals from the TDHF and frozen TDHF simulation results.
Without EHI, only electrons starting from $k_0 \in [-\frac{\pi}{a}, -\frac{\pi}{a}-\min(A(t))] = [-0.87, -0.54]$ and $k_0 \in [\frac{\pi}{a} - \max(A(t)), \frac{\pi}{a}] = [0.51, 0.87]$ climb up to the first CB by tunneling at $k = \pm \frac{\pi}{a}$ [Fig.~\ref{fig:electron final population}(a)].
Under the effect of EHI [Fig.~\ref{fig:electron final population}(b)], on the other hand, electrons occupy a much broader range of $k_0$ in the first CB, and moreover, those initially at $k_0 \in [-\max(A(t)), -\min(A(t))] = [-0.36, 0.33]$ [the arrow in Fig.~\ref{fig:electron final population}(b)] are promoted to the second CB.

\begin{figure}[t]
 \centering
 \includegraphics[width=0.98\linewidth]{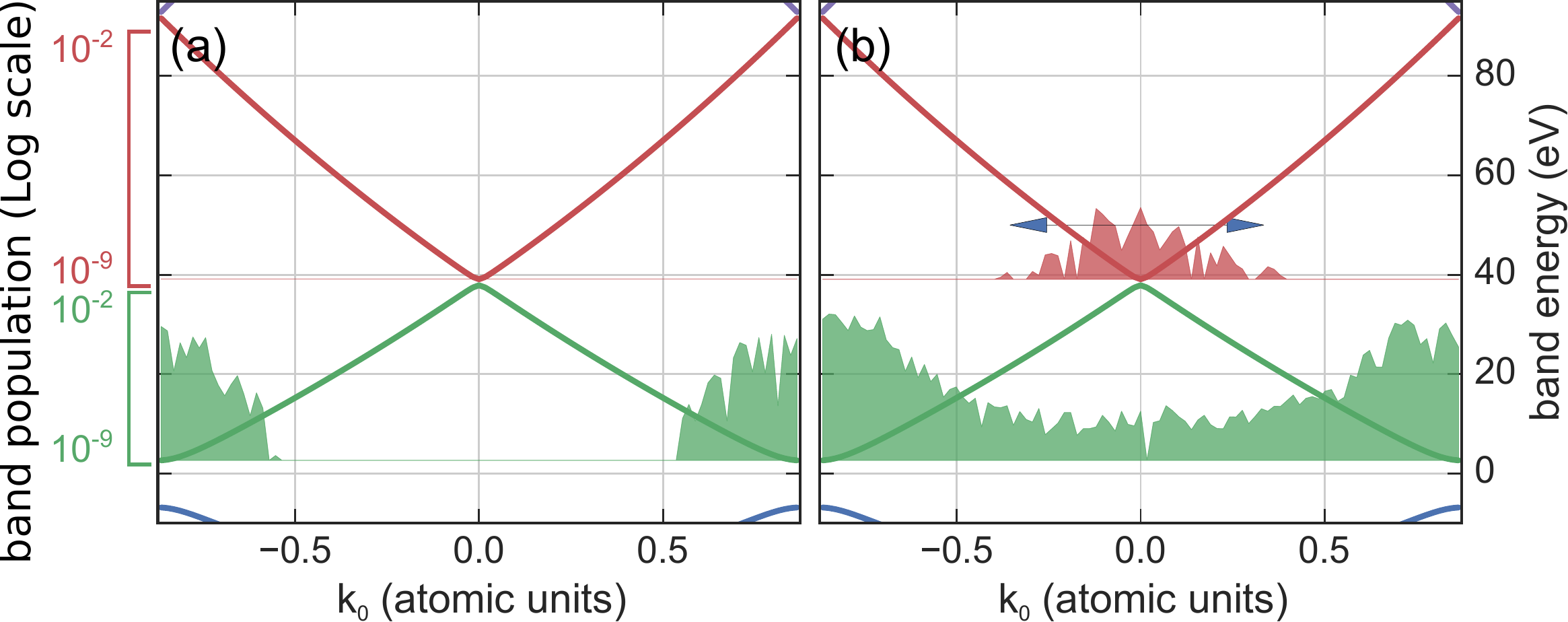}
 \caption{
 The final electron population of the first (green) and second (red) CBs projected onto the ground state Bloch orbitals from (a) frozen TDHF and (b) full TDHF simulations for $E_0 = 2.60$ V/nm. The horizontal arrow in (b) represents the range $k_0\in[-\max(A(t)), -\min(A(t))]=[-0.36, 0.33]$.
 }
 \label{fig:electron final population}
\end{figure}

In summary, we have investigated the effects of electron-hole interaction on HHG from solid-state materials based on TDHF simulations for a 1D model system.
We have found that, besides an exciton peak at low intensity, a second plateau appears at laser intensities much lower than expected from the independent electron approximation.
Using the Houston-basis expansion, we have identified it originating from the hauling-up effect due to EHI, where interband polarization, once created at and near an MBG, is capable of exciting electrons distant in the momentum space via Coulomb potential. It allows those electrons that cannot reach the MBG by the intraband displacement to climb up from the VB to the first CB.
This effect can be taken into account in the trajectory analysis to well reproduce the temporal structure of HHG extracted from the TDHF results.
If we shift our eyes back to the gas-phase HHG, the influence of the Coulomb potential from the parent ion, neglected in the strong-field approximation \cite{Lewenstein1994}, may somewhat correspond to that of EHI. 
However, it hardly affects qualitative features of harmonic spectra. 
Our results suggest that solid-state HHG involves much more complicated mechanisms than its gas-phase counterpart, and, therefore, offers even richer information on ultrafast many-body correlation dynamics in solid materials.

This research was supported in part by Grants-in-Aid for Scientific Research (No.~25286064, 26390076, 26600111, 16H03881, and 17K05070) from Japan Society for the Promotion of Science (JSPS), and also by the Photon Frontier Network Program of the Ministry of Education, Culture, Sports, Science and Technology (MEXT) of Japan. This research was also partially supported by the Center of Innovation Program from the Japan Science and Technology Agency, JST, by CREST (Grant No. JPMJCR15N1), JST, and by MEXT as ``Exploratory Challenge on Post-K computer". T.I. was supported by a JSPS Research Fellowship.

\bibliographystyle{apsrev4-1}
\bibliography{reference}

\pagebreak
\widetext
\begin{center}
\textbf{\large Supplementary material: Particle-hole interaction effects on strong-field driven electron dynamics in solids from time-dependent Hartree-Fock theory}
\end{center}
%


\renewcommand{\theequation}{S\arabic{equation}}
\renewcommand{\thefigure}{S\arabic{figure}}
\renewcommand{\bibnumfmt}[1]{[S#1]}
\renewcommand{\citenumfont}[1]{S#1}

\section{Interelectronic interaction operator in the time-dependent Hartree-Fock theory} 
The operator $\hat{w}[\rho]$ in Eq.~(1) describes the contribution from the interelectronic Coulomb interactions, composed of the Coulomb and exchange terms,
\begin{equation}
 w[\rho]({\bm x}, {\bm x}^{\prime}) = \int d{\bm y} \rho({\bm y}, {\bm y}) v({\bm x}, {\bm y}) \delta({\bm x}, {\bm x}^{\prime}) - \frac{1}{2} \rho({\bm x}, {\bm x}^{\prime}) v({\bm x}, {\bm x}^{\prime}),
\end{equation}
where $v({\bm x}, {\bm x}^{\prime}) = v(|{\bm x} - {\bm x}'|)$ is the Coulomb potential.
This operator acts on an orbital $\psi_{b {\bm k}_0}({\bm x}, t)$ as,
\begin{equation}
 \left[\hat{w}[\rho] \psi_{b {\bm k}_0}(t)\right]({\bm x}) = \int w[\rho]({\bm x}, {\bm x'}) \psi_{b {\bm k}_0}({\bm x'}, t) d{\bm x'}.
\end{equation}

\section{Linear response}
Linear response of a system is characterized by the linear optical conductivity
\begin{equation}
 \sigma(\omega) = \frac{j(\omega)}{E(\omega)} = \frac{\int dt e^{i\omega t} j(t)}{\int dt e^{i \omega t} E(t)}.\label{eq:linear optical conductivity}
\end{equation}
The linear optical conductivity is related to the dielectric function $\epsilon(\omega) = 1 + \frac{4 \pi i}{\omega} \sigma(\omega)$, and thus its real part corresponds to absorption.
Eq.~(\ref{eq:linear optical conductivity}) enables us to obtain the linear response of the system from real-time simulation with sufficiently weak electric field $E(t)$.
One of the convenient choices of $E(t)$ is an impulsive field $E(t) = E_0 \delta(t)$ ($|E_0| \ll 1$), which corresponds to the vector potential whose waveform is a step function $A(t) = - \int^{t}_{-\infty} dt' E(t') = - E_0 \Theta(t)$.

\begin{figure}[ht]
 \centering
 \includegraphics[width=0.5\linewidth]{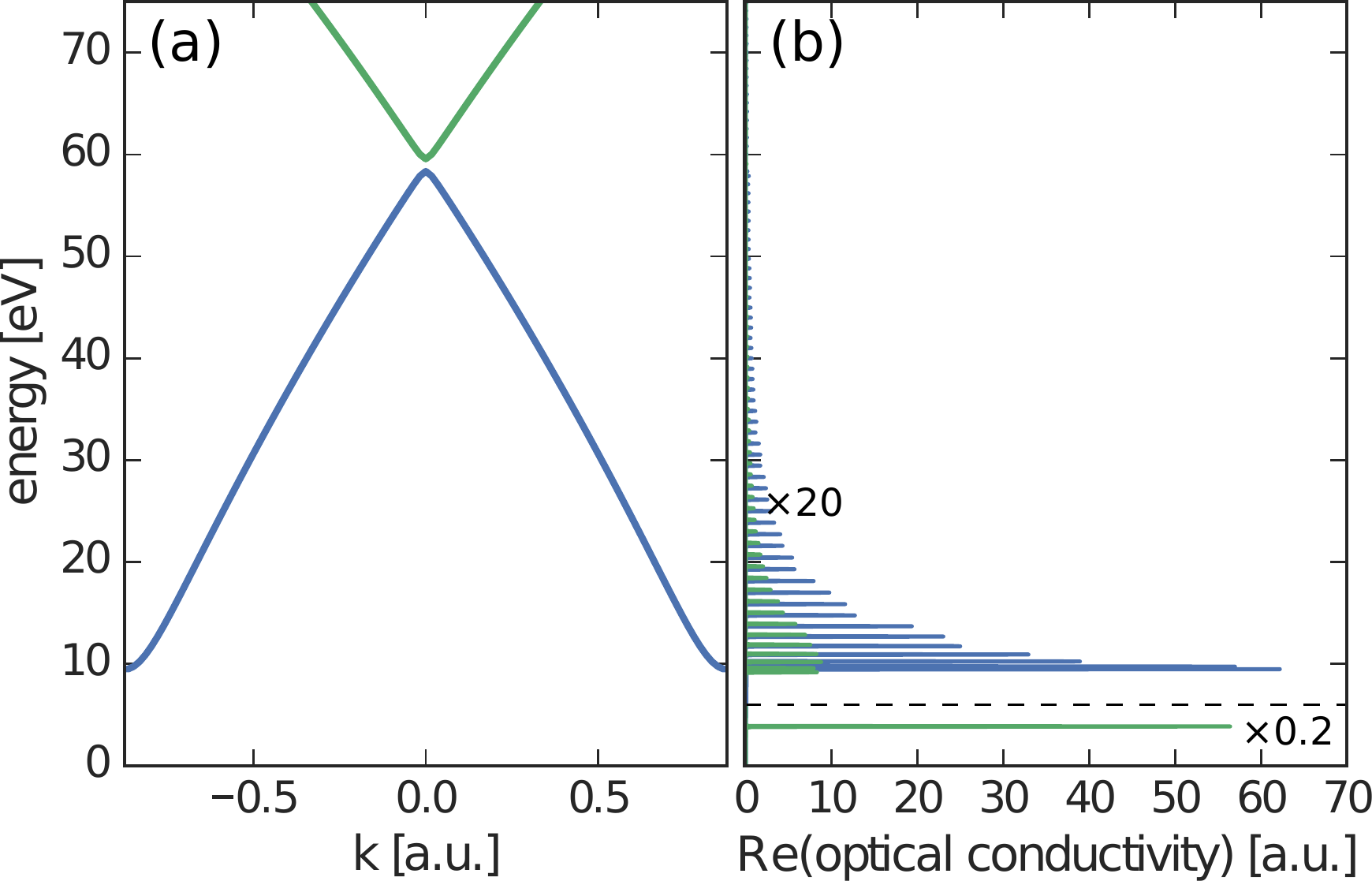}
 \caption{
 The band structure of the model hydrogen chain and its linear response.
 (a) The conduction band energy difference from the valence band $\Delta \varepsilon_{mn}(k) = \varepsilon_{m, k}-\varepsilon_{n, k}$ ($m = 1, 2$ and $n=0$).
 (b) The real part of the linear optical conductivity obtained from the real-time (green) TDHF and (blue) frozen TDHF simulations.
 As the TDHF conductivity has a strong exciton peak at $3.8$ eV, its below-gap component ($\le 6$ eV) is scaled by a factor of $0.2$ while its above-gap component is magnified by a factor of $20$.
 }
 \label{fig:linres}
\end{figure}

We show the real part of the linear conductivity obtained from TDHF and frozen TDHF simulation for the model hydrogen chain presented in the text in Fig.~\ref{fig:linres}.
We see a sharp exciton peak at $3.8$ eV well below the gap energy $9.5$ eV in the full TDHF case, while it is absent in the frozen TDHF result.

\section{Derivation of the TDHF equations in Houston basis}
\begin{figure}[ht]
 \centering
 \includegraphics[width=0.5\linewidth]{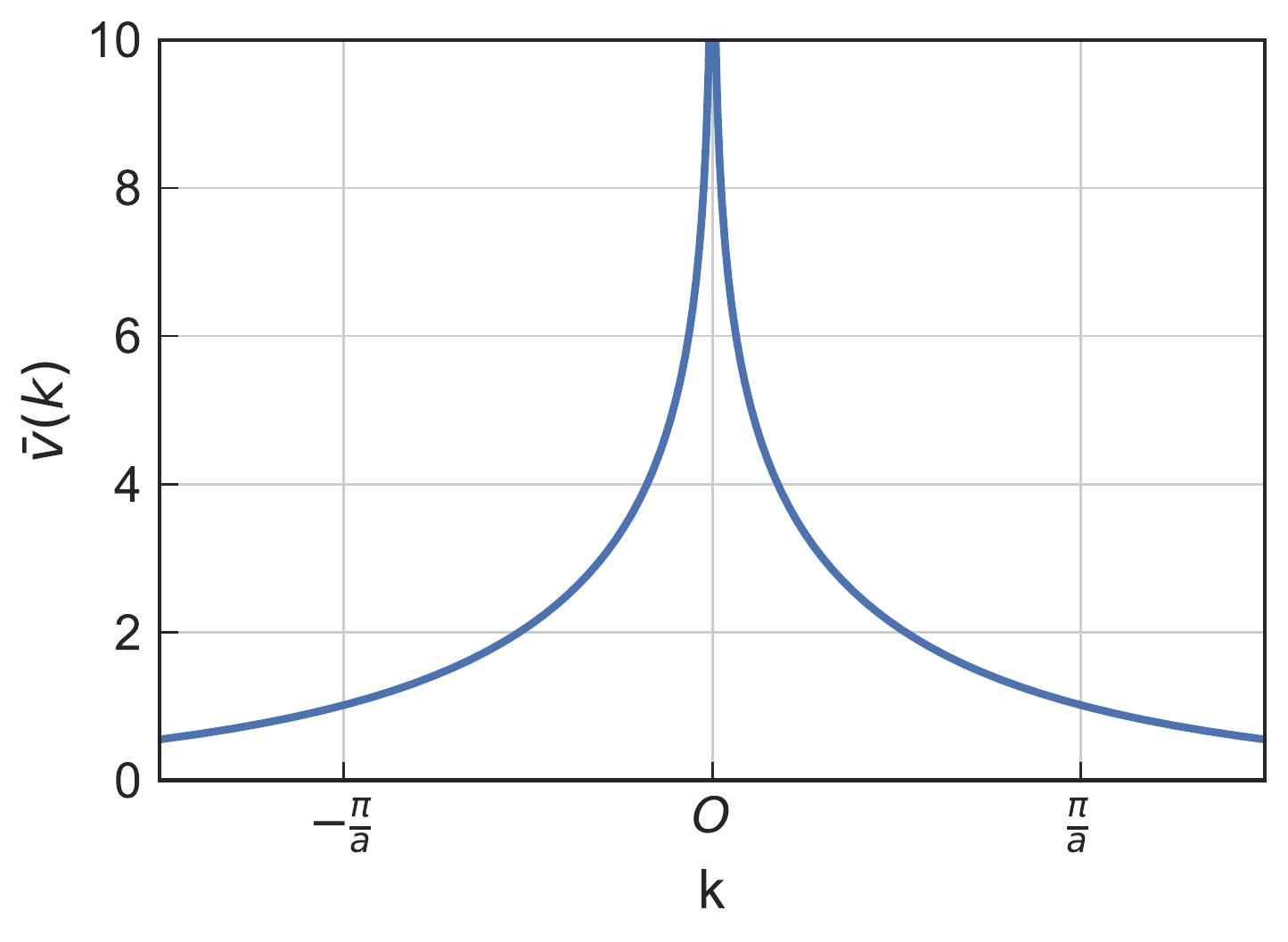}
 \caption{The Fourier transform of the soft Coulomb potential.}
 \label{fig:FT soft Coulomb}
\end{figure}

Here we present a derivation of Eq.~(6), i.e., the TDHF equations in the Houston basis.
Substituting Eq.~(5) into Eq.~(1) yields the coupled equations for complex amplitudes $\alpha_{k_0}^m(t)$ expressing electron {\it interband} dynamics,
\begin{equation}
 i \dot{\alpha}_{bk_0}^{m}(t) = \sum_{n} \alpha_{bk_0}^{n}(t) e^{i\int \varepsilon_{mn}[k(t')] dt'} \left( d_{k(t)}^{mn} E(t) + \bra{\tilde{\phi}_{mk_0}} \hat{w}[\delta\rho(t)] \ket{\tilde{\phi}_{nk_0}}\right),\label{eq:EOS for Houston probability amplitude}
\end{equation}
where $d_{k(t)}^{mn} = i\braket{u_{mk(t)} | \nabla_k u_{nk(t)}}$ with $k(t) = k_0 + A(t)$.
Here $u_{mk}(x)$ represents the lattice periodic part of the initial Bloch function, i.e., $\phi_{mk}(x) = e^{i k x}u_{mk}(x)$.
We calculate the matrix element of the interelectronic operator:
\begin{align}
 &\bra{\tilde{\phi}_{mk_0}} \hat{w}[\delta\rho] \ket{\tilde{\phi}_{nk_0}} \nonumber \\
 & = \iint dx dx' \tilde{\phi}_{mk_0}^{\ast}(x) \delta\rho(x', x') v(x, x') \tilde{\phi}_{nk_0}(x)
  - \frac{1}{2} \iint dx dx' \tilde{\phi}_{mk_0}^{\ast}(x) \delta\rho(x, x') v(x, x') \tilde{\phi}_{nk_0}(x') \nonumber \\
 & = \sum_{Gq} \bar{v}(G+q) \bra{\tilde{\phi}_{mk_0}} e^{i(G+q)x} \ket{\tilde{\phi}_{nk_0}} \Tr[\delta\rho e^{-i (G+q)x}]
 -\frac{1}{2} \sum_{Gq} \bar{v}(G+q) \bra{\tilde{\phi}_{mk_0}} e^{i(G+q)x} \delta \rho e^{-i(G+q)x}\ket{\tilde{\phi}_{nk_0}},\label{eq:matrix element of Coulomb operator}
\end{align}
where $\bar{v}(k)$ is the Fourier transform of the soft Coulomb potential $v(x, x')$, or
\begin{equation}
 v(x, x') = v(x - x') = \sum_{Gq} \bar{v}(G+q)e^{i(G+q)(x-x')}.
\end{equation}
Here $G$ is the reciprocal lattice vector: $G = 0, \pm 2\pi/a, \pm 4\pi/a \dots$, and $q$ takes the values within the Brillouin zone: $q \in [-\pi/a, \pi/a]$.
$\bar{v}(k)$ has an analytic form,
\begin{equation}
 \bar{v}(k) = \int dx \frac{e^{ikx}}{\sqrt{x^2 + 1}} = 2K_0(|k|),
\end{equation}
where $K_n(z)$ is the $n$-th modified Bessel function of the second kind (Fig.~\ref{fig:FT soft Coulomb}).

The first term in Eq.~(\ref{eq:matrix element of Coulomb operator}), which originates in the direct operator, can be transformed as 
\begin{align}
 &\sum_{Gq} \bar{v}(G+q) \bra{\tilde{\phi}_{mk_0}} e^{i(G+q)x} \ket{\tilde{\phi}_{nk_0}} \Tr[\delta\rho e^{-i (G+q)x}] \nonumber \\
 &= \sum_{G} \bar{v}(G) \bra{u_{mk(t)}} e^{iGx} \ket{u_{nk(t)}} \Tr[\delta\rho e^{-iGx}] \nonumber \\
 & = 2\sum_{G} \bar{v}(G) \gamma_{k(t), k(t)}^{mn}(G) \sum_{b' \in {\rm VB}, q}\left( \sum_{ij} D_{b'k(t)-q}^{ij}(t)\gamma_{k(t)-q, k(t)-q}^{ij\ast}(G) - \gamma_{k(t)-q, k(t)-q}^{b'b'\ast}(G) \right),
\end{align}
where we define
\begin{equation}
 \gamma_{kk'}^{mn}(G) = \bra{u_{mk}} e^{iGx} \ket{u_{nk'}},
\end{equation}
and
\begin{equation}
 D_{bk(t)}^{ij}(t) = \alpha_{bk_0}^{i}(t) \alpha_{bk_0}^{j\ast}(t) e^{-i\int \varepsilon_{ij}[k(t')]dt'}.
\end{equation}

The second term in Eq.~(\ref{eq:matrix element of Coulomb operator}), which stems from the exchange operator, is transformed as
\begin{align}
 &-\frac{1}{2} \sum_{Gq} \bar{v}(G+q) \bra{\tilde{\phi}_{mk_0}} e^{i(G+q)x} \delta \rho e^{-i(G+q)x}\ket{\tilde{\phi}_{nk_0}} \nonumber \\
 &= -\sum_{Gq} \bar{v}(G+q) \sum_{b' \in {\rm VB}}\left(\sum_{ij}D_{b'k(t)-q}^{ij}(t) \gamma_{k(t), k(t)-q}^{mi}(G) \gamma_{k(t), k(t)-q}^{nj\ast}(G) - \gamma_{k(t), k(t)-q}^{mb'}(G) \gamma_{k(t), k(t)-q}^{nb'\ast}(G) \right).
\end{align}
Therefore, the equation of motion for probability amplitude Eq.~(\ref{eq:EOS for Houston probability amplitude}) becomes
\begin{align}
 i &\dot{\alpha}_{bk_0}^{m}(t) = \sum_{n} \alpha_{bk_0}^{n}(t) e^{i\int \varepsilon_{mn}[k(t')] dt'} \Biggl[ d_{k(t)}^{mn} E(t) \nonumber \\
 &\quad +2\sum_{Gq} \bar{v}(G) \gamma_{k(t), k(t)}^{mn}(G) \sum_{b' \in {\rm VB}}\left( \sum_{ij} D_{b'k(t)-q}^{ij}(t)\gamma_{k(t)-q, k(t)-q}^{ij\ast}(G) - \gamma_{k(t)-q, k(t)-q}^{b'b'\ast}(G) \right) \nonumber \\
 &-\sum_{Gq} \bar{v}(G+q) \sum_{b' \in {\rm VB}}\left(\sum_{ij}D_{b'k(t)-q}^{ij}(t) \gamma_{k(t), k(t)-q}^{mi}(G) \gamma_{k(t), k(t)-q}^{nj\ast}(G) - \gamma_{k(t), k(t)-q}^{mb'}(G) \gamma_{k(t), k(t)-q}^{nb'\ast}(G) \right)
\Biggr].\label{eq:general EOM}
\end{align}

Moreover, we adopt some simplification and approximation to obtain a physical insight from Eq.~(\ref{eq:general EOM}).
First, we assume a single-VB system and omit the label $b'$ hereafter.
Second, we ignore the Coulomb potential $\bar{v}(k)$ outside the Brillouin zone, i.e., we assume $\bar{v}(k) = 0$ for $|k| > \frac{\pi}{a}$.
This assumption leads to the vanishing second term in Eq.~(\ref{eq:general EOM}) because $\Tr[\delta\rho] = 0$.
Third, we assume that $\gamma_{k, k'}^{mn}(0) = \braket{u_{mk} | u_{nk'}} \approx \delta_{nm}$.
Based on these assumptions, Eq.~(\ref{eq:general EOM}) can be simplified into Eq.~(6),
\begin{equation}
 i\frac{d}{dt} \alpha_{k_0}^m(t) \approx \sum_{n} \alpha_{k_0}^{n}(t) e^{i\int_0^t \varepsilon_{mn}[k(t^{\prime})]dt^{\prime}} \left[ E(t) d_{k(t)}^{mn} - \sum_{q \in {\rm BZ}} \bar{v}(-q) D_{k(t)+q}^{mn}(t) \right].\label{eq:EOM for Houston function expansion2}
\end{equation}



\end{document}